\documentclass[a4,12pt,epsfig,graphicx]{article}
\usepackage{pstricks}
\usepackage{epsfig}
\usepackage{graphicx}
\usepackage[latin2]{inputenc}
\usepackage{epic}
\usepackage{latexsym}
\usepackage{amssymb}

\begin{document}


\pagestyle{empty}
\setcounter{page}{0}
{\normalsize\sf
\vskip 3mm
\rightline{IFT/02/17}
\vskip 3mm
\rm\rightline{June 2002}
}
\vskip 2cm
\begin{center}
{\LARGE {Racetrack models 
in theories from extra dimensions}}
\vspace*{5mm} \vspace*{1cm} 
\end{center}
\vspace*{5mm} \noindent
\vskip 0.5cm
\centerline{\bf Rafa{\l} Ciesielski and Zygmunt Lalak}
\vskip 1cm
\centerline{\em Institute of Theoretical Physics }
\centerline{\em University of Warsaw, Poland }
\vskip 2cm
\centerline{\bf Abstract}
\vskip .5cm
\noindent We have investigated moduli stabilization leading to hierarchical supersymmetry breakdown 
in racetrack models with two moduli fields simultaneously present in the effective racetrack superpotential. We have shown that stabilization of moduli occurs when 
a shift symmetry of the moduli space becomes gauged. This gauging results in a $D$-term contribution to the scalar potential that depends on moduli scalars. 
To break supersymmetry at a minimum created this way in the case where only a single combination of moduli is present in the superpotential one needs supergravity corrections. 
If the superpotential depends on two independent combinations of moduli, supersymmetry is broken by non-vanishing $F$-terms without supergravity terms. Some of the minima that we see correspond to a non-vanishing expectation value of the blowing-up modulus in the case of type IIB orientifold models. 
We point out that the mass of the gauge boson associated with gauged shift symmetry becomes naturally light in warped compactifications.\\

\vskip1cm
\begin{flushleft}   
June 2002 
\end{flushleft}

\def\C{{ C\scriptstyle{\!\!\!\!|} }} 
\def\D{{D}} 
\def\F{{F}} 
\def\T{{T}} 
\def\t{{\mathcal{T}}} 
\def\S{{S}} 
\def\G{{\grave{\mathcal{G}}}} 
\def\W{{W}} 
\def\V{{V}} 
\def\B{{ \kappa_{c.c.} }} 
\def\b{{\Upsilon}} 
\def\K{{K}} 
\def\m{{ \hat{m} }} %
\def\0{{ \cal O }} %
\def\f{{ \cal F }} %
\def\h{{ h }} %
\def\mM{{ \cal M }}
\def\SU{{ SU(N) }}
\def\U{{ U(1) }} 
\def\UX{{ U(1)_X}}
\def\b0{{ \beta_0}}
\def\l{{ \Lambda }} 
\def\L2{{ \chi }}       
\def\fl{{  f_{\l} }}
\def\fx{{  f_{\L2} }}
\def\cl{{  c_{1} }}
\def\cx{{  c_{2} }}
\def\c{{  c }}
\def\kl{{  k_{\l} }}
\def\kx{{  k_{\L2} }}
\def\bl{{  \widetilde{b_1} }} 
\def\bx{{  \widetilde{b_2} }} 
\def\gama{{ \gamma }}
\def\P{{ M_P }}
\def\Nc{{  N_c  }}
\def\Nf{{  N_f  }}
\def\t{{  t  }}
\def\z{{  Z  }}
\def\s{{  S  }}

\setcounter{page}{1} \pagestyle{plain}
\section{Introduction}

Supersymmetry provides a technically natural and rich in physical implications 
method of controlling the hierarchy of mass scales in four-dimensional
field theories, to all orders 
in perturbative calculations.
However, supersymmetry must be broken at low energies to account properly for observable phenomenology. The most economic way of doing this is to couple the gauge and matter Lagrangian to gravity in the locally supersymmetric manner and to break local supersymmetry spontaneously. 
First, one avoids the appearance of the massless fermion -- the Goldstino becomes the \mbox{spin $1/2$- component} of the massive gravitino. Second, weak gravitational couplings account for the 
natural suppression of the communication of the effects of the supersymmetry breakdown to the 
observable sector. In the flat limit this procedure gives rise to the Lagrangian with 
explicitly but softly broken global supersymmetry. 
On the other hand in theories borne in higher dimensions there exist naturally light fields, 
which are neutral under gauge interactions and form flat or run-away directions in the field space. Such moduli fields arise very naturally in theories with extra dimensions as extra-components of 
the higher-dimensional metric tensor, form fields and gravitini.  After compactification such fields 
may contribute to the four-dimensional cosmological constant, through their potentials and transverse gradients. Hence, what one really requires from the successful scenario of supersymmetry breakdown 
is not only generation of 1 TeV mass splittings in observable supermultiplets, but also 
stabilization of moduli while achieving a nearly vanishing four-dimensional cosmological constant. 

Among various proposals heading in this direction a particularly simple one is the idea of using
several gaugino condensates, i.e. the racetrack models 
\cite{Krasnikov:jj},\cite{Dixon:1990ds},\cite{clmr}.
Technically it reduces down to generating a superpotential containing several components that 
depend exponentially on moduli scalars. As demonstrated over the years
such exponential contributions may arise not only from strongly
interacting gauge sectors, but also from nontrivial  warp 
factors along transverse dimensions or/and brane solutions of higher dimensional supergravity and string theories. Another ingredient coming from type I and type II orientifold theories are additional, twisted, moduli 
\cite{Aldazabal:1998mr}. These moduli enter, along the untwisted dilaton (and sometimes radion) the kinetic functions of some nonabelian gauge groups. Further to this, in type II B orientifold models there appear 
anomalous $U(1)$ gauge factors, whose effective \mbox{Fayet-Iliopoulos} terms are proportional to the background values of the twisted moduli. Hence, these moduli obtain an additional contribution to their potential through the anomalous D-terms. 
The presence of the radion in the gauge kinetic functions has been found long ago in the case of weakly coupled heterotic superstring \cite{Itoyama:1985ay},\cite{Ibanez:1986xy}, then in the case of the strongly coupled heterotic superstring \cite{Banks:1996ss}, and recently in warped five-dimensional brane models \cite{Falkowski:2001sq}. 

In what follows we shall investigate to what extent one can achieve successful supersymmetry breakdown and moduli stabilization with the use of the generalized racetrack models 
in four-dimensional supergravities enhanced by the extra-dimensional features discussed above. 


\section{Moduli dependent Fayet-Iliopoulos terms}
First, let us summarize briefly the features of the four-dimensional models with
field-dependent FI terms that are relevant for this note.  
Basically, such terms can be understood as D-terms arising upon
gauging of global translations along direction of certain moduli
fields. Let us call a representative modulus $Z$, and assume that its
K\"ahler function depends only on $Z + \bar{Z}$, $K=K(Z + \bar{Z})$. 
Then the shift \mbox{$Z \rightarrow Z -i \delta /2\, \Lambda$} is the isometry of the
associated K\"ahler manifold generated by the Killing vector $X=-i
\delta /2$. To gauge this isometry we introduce in the supersymmetric
way the massless vector
field $A_\mu$ which enters the covariant derivative $D_\mu =
\partial_\mu + i g \delta /2 A_\mu$, where $g$ is the gauge
coupling constant. Due to supersymmetry there appears the prepotential $D$,
fulfilling the Killing equation $ K_{\z \bar{\z}} \bar{X} = i \partial
D/\partial \z$. This prepotential generates the scalar potential for
$\z$, $V= \frac{1}{2} g^2 D^2$, and plays the role of the field
dependent Fayet-Iliopoulos term $\xi^2$, $ \xi^2 = D$. It is easy to
see that, up to a constant, the $D$/FI term associated with the
Killing vector $X$ is $D= \delta /2 \partial K / \partial \bar{\z}$. 
In addition, the covariantized kinetic term for the scalar $\z$ gives
rise to the mass term for the vector boson, 
\begin{equation}
m_{A}^2 = \frac{M_{P}^2 g^2
  \delta^2}{2} K_{\z \bar{\z}},
\end{equation} 
where we have put explicitly the
  4d Planck scale. In general, when the gauge coupling $g$ is field
  dependent, $g=g(z)$, the local shift of $\z$ is anomalous, and one
  needs to introduce other fields charged under the gauge symmetry 
to cancel the anomaly. This issue has been discussed at length 
in a number of papers and in what follows we assume that such a
compensation is possible whenever we need it (but see \cite{Ibanez:1999pw},\cite{Lalak:1999ex}). 
The well known case of a field-dependent D-term is the one of the
heterotic string, where one gauges the imaginary shift of the
dilaton. There $K=-\log (S + \bar{S})$,\mbox{ $D=(M_{P}^2 \delta )/( 2
(S + \bar{S}))$}, and the mass of the gauge boson is naturally of the
order of the Planck scale. The case of type IIB orientifolds
corresponds to $K= \frac{1}{2} (M + \bar{M})^2$ (see \cite{P} and
references therein). This gives $D= \delta /2 (M + \bar{M})$. 
An interesting possibility is offered by warped compactifications 
\cite{Falkowski:2001sq},\cite{Lalak:2001dv}. 
There $K=-3 \log (f(T + \bar{T}))$, where the function $f$
reflects the vacuum configuration of the warp factor along the
extra-dimensions. To be specific let us take the case of the Randall-Sundrum
model, 
$f=\beta ( 1 - e^{-(T + \bar{T})})$. This results in the
Fayet-Iliopoulos term 
\begin{equation}
D= - \frac{\delta}{2} M_{P}^2 \frac{e^{-(T +
    \bar{T})}}{1 -e^{-(T + \bar{T})}},
\end{equation} 
and the mass of the gauge
boson is 
\begin{equation}
m_{A}^2 = \frac{3 g^2 \delta^2}{2} M_{P}^2 \frac{e^{-(T +
    \bar{T})}}{1 - e^{-(T + \bar{T})}}.
\end{equation}     
In the case of the
Randall-Sundrum model the warp factor can easily be taken in such a
way, that the mass scale one the warped brane, $e^{-(T + \bar{T})}
M_{P}^2$ becomes $(1 \; {\rm TeV})^2$. Then of course the mass of the
gauge boson, and the field dependent FI term are also of the order $(1
\; {\rm TeV})^2$. Hence, in the warped case the gauge boson associated
with the `anomalous' $U(1)$ group may be naturally light. Of course
the microscopic picture is likely to somewhat more complicated, as for
instance one expect generation of the moduli dependent superpotential 
in the warped models, which may break the imaginary shift symmetry. 

\section{Moduli stabilization}

In what follows we shall concentrate on a model resemblig the
structure of  models derived from type IIB orientifolds, with two
types of moduli: the dilaton $S$ and the analog of a twisted modulus $M$.  
We shall assume two $SU(N_i), \; i=1,2$ gauge group with gauge
dependent kinetic functions $f_i = S + c_i M$.  
Condensation of the gaugini of the two $SU(N)$s give rise to the
exponential superpotential for the moduli:
 \begin{eqnarray}\label{W}
W(S,M) &=&  
        \P^3 \Bigg(\alpha e^{\frac{-24\pi^2}{b_1} \Big( \frac{S + \cl M}{\P} \Big)} 
                  + \beta e^{\frac{-24\pi^2}{b_2} \Big(\frac{ S + \cx M}{\P} \Big)}\Bigg)
 \end{eqnarray}
where the $b_i$ are one-loop beta function coefficients normalized through 
\mbox{$\;\beta (g_i) = - \frac{b_i}{(4\pi)^2} g_{i}^3\;\;$},\\
and $\alpha$, $\beta$ are 
model dependent parameters.

We are interested in the question of stabilization of moduli and
hierarchical supersymmetry breaking in such a setup, in the cases of
globally and locally supersymmetric models. In fact, it would be
natural to start with a single exponential term in the superpotential,
and to play with possible variation of the K\"ahler function to
stabilize the moduli. However, within the restricted class of K\"ahler
potentials we consider we were unable to find a minimum of the potential
neither in the globally supersymmetric nor in the supergravity version
of the model\footnote{In \cite{AbelServant:2001ab} it is argued that a single condensate 
may lead to stabilization in a suffciently complicated type IIB model. 
Here we do not want to restrict ourselves by assuming special features of the 
moduli potential, like for instance a modular invariannce. We would like 
the stabilization to take place independently of particular details of a model.}. Moreover, it turns out to be difficult to find 
a minimum of the potential even with two exponential terms 
(`two condensates'). That this would be the case one could foresee on
the basis of the negative result in the case $c_1=-c_2$ considered
in \cite{Lalak:1999yq}
 in the context of the Horava-Witten model \cite{Horava:1996ma}. 
The strategy of the search for reasonable minima with hierarchical
susy breaking is analogous to that of \cite{Lalak:1999yq}: first we try to find 
minima (with unbroken susy) of the globally supersymmetric Lagrangian,
and then we switch on gravity-induced corrections. In addition, we
shall check the sensitivity of the minimization with respect to a
deformation of the K\"ahler function for the $M$-modulus, and we shall
switch on the $M$-dependent Fayet-Iliopoulos term. 

We start with the case that can be followed analytically.    
The first simplifying assumption is that the kinetic functions are the
same, $  f_1= f_2$. 
This means that $\cl =\cx = c $. Then we perform a holomorphic redefinition
of variables $S,\;M$:
 \begin{eqnarray}\label{f_1}
 f_1 = f_2 =  S +\c M  \equiv  \z .\\
 \end{eqnarray}
In new variables the K\"ahler potential takes the form
 \begin{eqnarray} \label{kahler_1}   
   \K (\s,\bar{\s},\z,\bar{\z}) &=& - \ln (\s+\bar{\s})+ \frac{1}{2 \c^2} (\z + \bar{\z} -S - \bar{S})^2
               + \frac{\gamma}{4 \c^4} (\z + \bar{\z} -S - \bar{S})^4 ,\nonumber \\
\end{eqnarray}
where $\gamma$ is a new, presumably small, real parameter measuring
the departure from the quadratic K\"ahler potential for $M$. 
The superpotential becomes
 \begin{eqnarray}\label{W_1}
W (\z)&= &\alpha e^{-\frac{\z}{\bl}}+ \beta  e^{-\frac{\z}{\bx}}  
 \end{eqnarray}
where $ \widetilde{b_i}=\frac{b_i}{24\pi^2}$,
and the inverse K\"ahler metric is
\begin{equation}\label{Metrics}
g^{i \bar{j}} = (\s+\bar{\s})^2
\left(
\begin{array}{cc}
\frac{1}{(\s+\bar{\s})^2}\frac{\c^4}{\c^2 + 3 \gamma(\z + \bar{\z} -\s -\bar{\s})^2 } +1 & 1\\
 1 & 1\\
\end{array}
\right).
\end{equation}
The scalar potential contains, in the global limit,
 \begin{eqnarray}\label{V_1}
V (\s,\bar{\s},\z,\bar{\z}) =g^{\z \bar{\z}} 
   \Big|\frac{\partial W(\z)}{\partial \z }\Big|^2 
    + \frac{1}{2} g^2 D^2, 
 \end{eqnarray}
where $ D $ is the $U(1)$ $D$-term\footnote{We assume that we are restricted to
the flat directions of non-abelian $D$-terms.} 
 \begin{eqnarray}\label{D_1}
D = \frac{1}{2 c^2} ( \z + \bar{\z}- \s - \bar{\s} + \frac{\gamma}{c^2} (\z + \bar{\z}- \s - \bar{\s})^3).
 \end{eqnarray}
%
The scalar potential of the model in new variables $Z, \; S$ before gauging the imaginary translation takes the form
 \begin{eqnarray}
V (\s,\bar{\s},\z,\bar{\z}) = 
 \bigg(\frac{\c^4}{\c^2 + 3 \gamma(\z + \bar{\z} -\s -\bar{\s})^2 } +(\s+\bar{\s})^2 \bigg) 
\Big|\frac{\partial W(\z)}{\partial \z }\Big|^2 .
 \end{eqnarray}
This potential has a flat direction, 
 which is located along $S$ for $Z$ fulfilling 
the condition for unbroken global supersymmetry $\partial W / \partial Z =0$: 
 \begin{eqnarray}\label{z_01}
  \frac{\partial W(\z)}{\partial \z }=0 
   & \Rightarrow &  
  \z_0 = \frac{\bl \bx}{\bx-\bl} \ln \bigg(-\frac{\beta \bl}{\alpha \bx}\bigg).
 \end{eqnarray}
Putting in the nonzero coefficient $\gamma$ of the quartic 
term in the K\"ahler potential doesn't change the situation. 

However, switching on the field-dependent Fayet-Iliopoulos term makes a difference. To simplify the reasoning and to make it somewhat model-independent let us replace the original superpotential by its expansion around the point $Z_0$ such that $\partial W / \partial Z |_{Z_0} =0$%
 \begin{eqnarray}\label{w_n}
W (\z) =  w + (\z-\z_0)^2 w_2 
 \end{eqnarray}
where $ w= W (\z_0)$ , 
and $ w_2=\frac{\partial^2 W(\z)}{\partial \z^2 }|_{\z_0}$.  
One readily finds a minimum at the point  
$ S = \z_0 $ and $\z  =\z_0$.
Hence the D-term contribution to the potential localizes $S$ with respect to $Z$, 
which in turns gets localized by the superpotential.

In the presence of the nonzero FI term there exists a minimum of the potential for any real  
$\gamma$, thus the presence of the quartic term in the K\"ahler 
potential for $M$ doesn't change the picture qualitatively. Hence, in the globally supersymmetric 
versions of the models that we consider, there appears a minimum of the scalar potential for 
the moduli fields, and this minimum preserves supersymmetry.

\section{Breakdown of supersymmetry}

To check whether the usual, and perhaps the most favourable, scenario
where supersymmetry breaking is triggered by gravitational corrections takes place 
we embed the model into the standard $N=1$ supergravity Lagrangian. 
Firstly,  we switch off the FI term.  
The  scalar potential takes the usual form  
 \begin{eqnarray}
V =e^{K} \Bigg\{ 
             \Big( \c^2 +(S + \bar{S})^2 \Big) \bigg| \frac{\partial W(\z)}{\partial \z} +\frac{\partial K}{\partial \z}W\bigg|^2 +
   \;\;\;\;\;\;\;\;\;\;\;\;\;\;\;\;\;\;\;\;\;\;\;\;\;\;\;\;\;\;\;\;\;\;\;\;\;\;\;\;\;\;\;\;\;\;\;\;\;\;\;\;\;\;\;\;\;\;\;\;\;\;\; \nonumber \\ \nonumber
             \Big( (S + \bar{S})^2 \bar{W}\frac{\partial K }{\partial \bar{S}}  \bigg(\frac{\partial W(\z)}{\partial \z} +\frac{\partial K}{\partial \z}W\bigg) +c.c. \Big) 
            + (S + \bar{S})^2 \bigg| W \frac{\partial K }{\partial S} \bigg|^2    -3\bigg| W\bigg|^2
            \Bigg\} \nonumber\\
 \end{eqnarray}
with  $K$ and $W$ given by  (\ref{kahler_1}) and (\ref{w_n}). \\ 
We are searching for a minimum which is close to the supersymmetric minimum of the 
globally supersymmetric Lagrangian, 
hence we expand the fields $S$ and $Z$ around that  
point: $Z=Z_0 (1 +\delta)$ and $S=Z_0 (1 + \epsilon)$. 
Expanding in $\delta$ and $\epsilon$ one obtains rather 
clumsy expressions, which however show that the flat direction that 
was present in the simplest version of the globally supersymmetric model turns into a run-away 
direction. For instance in the lowest order of the expansion in $\delta, \; \epsilon$ and in the limit 
$M_P \rightarrow \infty $ the expression for $\delta$ becomes 
 \begin{eqnarray}\label{delta_01}
\delta =  \frac{(2 S -Z_0) w}{2 Z_0 c^2 M_{P}^2 w_2};
\end{eqnarray}
this correction necessarily depends on the vacuum value of $S$, which signals the run-away 
behaviour.   

Again, one can see from these formulae, and check using numerical analysis, 
that there is no minimum 
created by the correction to the quadratic K\"ahler potential. 
The model exhibits the run-away behaviour in the $(S,Z)$ plane for any value of $\gamma$.
 
\begin{figure}
        \centering\epsfig{file=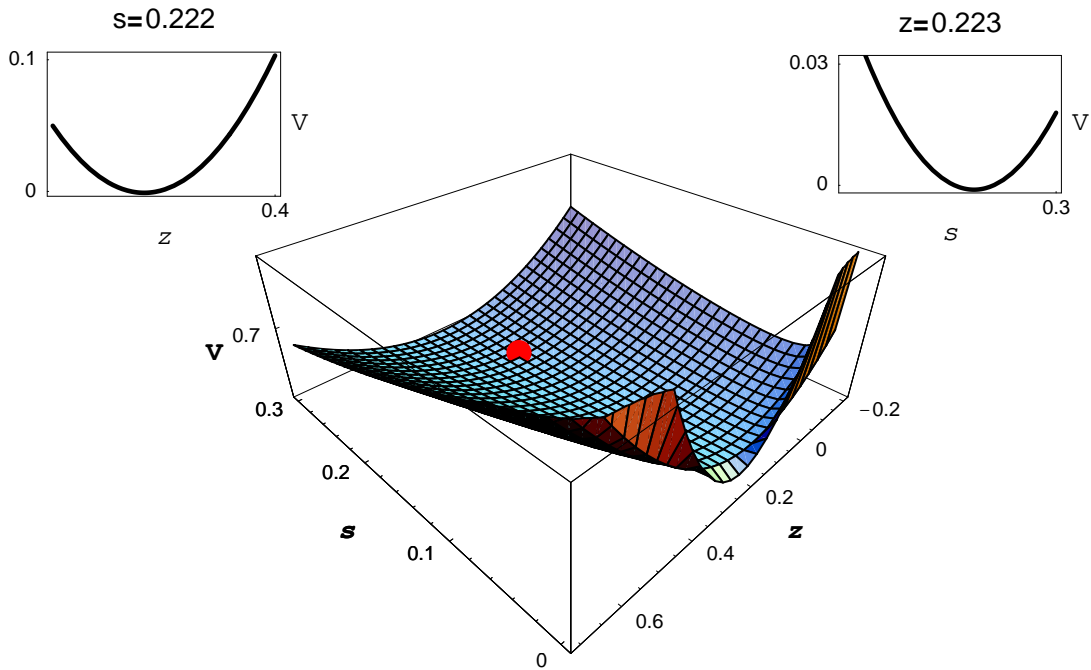,width=.9 \linewidth}
        \caption{The picture illustrates formation of a supersymmetry breaking minimum 
        in the supergravity scalar potential when superpotential depends on 
        a single combination of moduli, and the shift symmetry is gauged. 
        For better illustration somewhat unrealistic values of parameters
        were assumed:   
        $\gamma = 0 $,
         $w = .1, w2 = .01, \z_0 = 0.24, g^2 = .001, M_P = 50, 
              c = \sqrt{2} $.  \label{fig_SUGRA_D=/=0_gamma=0}}
     \end{figure}

Finally, we switch on the Fayet-Iliopoulos term. 
As expected, there appears a minimum,  and it is located at the point
\begin{eqnarray}
S= \z_0 \Big( 1 + \epsilon \Big), \;\; \z =  \z_0  \Big( 1 + \delta \Big),
\end{eqnarray}
where the exact lowest order solutions for $\epsilon$ and $\delta$ are given in the appendix, formulae 
(\ref{epsi}),(\ref{delt}). 
These expressions in the flat space limit $M_P \rightarrow \infty$ give 
\begin{eqnarray}
\epsilon = - \frac{1}{4 Z_{0}^2 c^2} \frac{w^2}{M_{P}^2 w_{2}^2},
\end{eqnarray}
and 
\begin{eqnarray}
\delta  =  \epsilon +  \frac{c^2 w }{8g\z_0^3}\frac{w}{M_P^3}.
\end{eqnarray}
Formation of this supersymmetry breaking minimum is illustrated in 
fig. \ref{fig_SUGRA_D=/=0_gamma=0}.

One can calculate the $F$-terms, and in the flat limit they are given by 
\begin{eqnarray}\label{F_z_SUGRA}
F^\z = F^S = -2 \sqrt{2} \frac{w}{M_P}  \left ( \frac{Z_0}{M_P} \right )^{3/2}. 
\end{eqnarray}
The vacuum value of the $D$-term is
 \begin{eqnarray}
D = \frac{1}{c^2} ( \z_0 (\delta-\epsilon) + \frac{4 \gamma}{c^2} {\z_0}^3 (\delta -\epsilon)^3);
 \end{eqnarray}
in the same approximation as the one used for the $F$-terms this gives
\begin{equation} 
D =  \frac{w}{8g\z_0^2} \frac{w}{M_P^2}. 
\end{equation}
Complete lowest order expressions for the $F$-terms are quoted in the appendix, formulae 
(\ref{Fz}),(\ref{Fs}). 
 
At this point we note that the dependence on $\gamma$ appears only in the third order 
in the expansion  in $\epsilon,\; \delta$,  as seen from the expressions for $F$-terms (\ref{Fz}),(\ref{Fs}), and that this dependence is irrelevant for the stabilization of moduli and for the size of the 
supersymmetry breaking effects. 

Using the above formulae for $F$-terms and $D$-terms one can compute the corrections to 
physical masses of matter scalars in the background corresponding to the minimum which we have found.
The contribution due to the $F$-terms is 
\begin{equation} 
\delta_{F} m^2 = \frac{8}{3 (\S + \bar{S})^2 } \left ( \frac{Z_0}{M_P} \right )^3 \left (\frac{w}{M_{P}^2} \right )^2,
\end{equation}
and the contribution due to the $D$-term looks as follows
\begin{equation}
\delta_{D} m^2 = g Q \frac{w^2}{4\z_0^2 M_P^2},
\end{equation}
where $Q$ is the $U(1)$ charge of a scalar field. 
One can see that gravitational suppression of the $D$-term contribution is milder, 
hence for $U(1)$-charged matter this contribution shall 
dominate\footnote{We are  assuming that only moduli take on non-zero vevs.}. 

\section{Superpotential dependent on two combinations of moduli}

We have seen that the presence of moduli dependent $D$-terms associated 
with a gauging of some noncompact symmetry of the moduli space leads to supersymmetry breaking 
stable vacua in generalized racetrack models. The assumption that has been made so far is 
that it is a single combination of moduli fields that enters the racetrack superpotential, i.e. 
that coefficients $c_i$ in the kinetic functions for both gauge groups are the same.  
Fortunately, it is possible to check that a mild splitting between $c_1$ and $c_2$ doesn't spoil
validity of our observations.

Let us set 
 \begin{eqnarray}\label{f_n1}
 f_1 =  S + c_1 M  \equiv  \z,   & & f_2 =  S + c_2 M \equiv S (1 -c_2 /c_1) + \z c_2/c_1, 
 \end{eqnarray}
where $  c_2 - c_1 \equiv \eta $ and $\eta \ll 1$.
The K\"ahler potential  takes the form (\ref{kahler_1}),  but the superpotential obtains a correction
\begin{eqnarray}\label{w_n_eta}
W (\z,\s) =  w + (\z-\z_0)^2 w_2 - \eta h (\z -\s), 
 \end{eqnarray}
where $h = \frac{\beta }{c_1\bx} e^{-\frac{\z}{\bx}} $ is approximated by a constant.
With this changes taken into account the scalar potential assumes the standard supersymmetric  
form, where  $K$ and $W$ are given by (\ref{kahler_1}) and (\ref{w_n_eta}), and the $D$-term is given
by (\ref{D_1}).
One finds a supersymmetry breaking minimum at the point  $\z  =\z_0(1+\delta_{split})$,
$ S = \z $, with the complete expression for $\delta_{split}$  given by the formula (\ref{delsp}) 
in the appendix.  
By inspection of the expressions for the $F$-terms one finds that this time it is impossible to 
have simultaneously $F^S=0$ and $F^Z=0$. Thus after the splitting of gauge kinetic functions 
there appears a supersymmetry breaking vacuum without introducing gravitational corrections. 
In this minimum $\langle D \rangle =0$. When one considers instead the full supergravity 
scalar potential, the situation becomes qualitatively very similar to the one described already 
in the case of two identical gauge kinetic functions. Numerical analysis of the complete model 
with two exponentials in the superpotential confirms above observations.

\section{Summary}

We have investigated moduli stabilization and supersymmetry breakdown 
in racetrack models with two moduli fields simultaneously present in the effective racetrack superpotential. We have demonstrated that stabilization of moduli occurs when 
a shift symmetry of the moduli space becomes gauged. This gauging results in a $D$-term contribution to the scalar potential that depends on moduli scalars. \\
To break supersymmetry at a minimum created this way in the case where only a single combination  of 
moduli is present in the superpotential one needs to switch on supergravity corrections. Then $F$-terms of all 
moduli as well as the expectation value of the $D$-term are non-zero, and it is the $D$-term contribution 
that dominates soft scalar masses. If the superpotential depends on two 
independent combinations of moduli, then supersymmetry is broken by non-vanishing 
$F$-terms even without supergravity corrections. Some of the minima that we see correspond 
to a non-vanishing expectation value of the blowing-up modulus (M) in the case of type IIB 
orientifold models. \\
In addition, we have pointed out, that the gauge boson associated with gauged shift symmetry becomes higgsed and its mass  may be naturally light in warped compactifications.

\vskip 1cm

\noindent{\large \bf Acknowledgments}

\vskip .3cm

\noindent
This work  was partially supported  by the EC Contract
HPRN-CT-2000-00152 for years 2000-2004, by the Polish State Committee for Scientific Research grant KBN 5 P03B 119 20 for years 2001-2002, and by POLONIUM 2002. 


\vspace{1.5cm}
\appendix
\noindent{\large \bf Appendix }
\setcounter{equation}{0} 
\renewcommand{\theequation}{A.\arabic{equation}}
\vspace{0.3cm}

{\footnotesize 

In the model with only a single combination of moduli present in the superpotential, and with gauged 
imaginary translations  there appears a minimum located at the point
\begin{eqnarray}
S= \z_0 \Big( 1 + \epsilon \Big), \;\; \z =  \z_0  \Big( 1 + \delta \Big),
\end{eqnarray}
with
\begin{eqnarray} \label{epsi}
\epsilon = \frac{w\left( c^2ww_2\left( w + c^2{M_P}^2w_2 \right)  + 
      2{M_P}^3\z_0\left( w - 4w_2{\z_0}^2 \right) g  \right) }{2c^2w^2w_2
     \left( w + c^2{M_P}^2w_2 + 6w_2{\z_0}^2 \right)  + 
    4{M_P}^3\z_0\left( w^2 + 2w_2\left( w - c^2{M_P}^2w_2 \right) {\z_0}^2 - 
       8{w_2}^2{\z_0}^4 \right) g }, \nonumber \\
\end{eqnarray}
and $\delta$ is given by
\begin{eqnarray} \label{delt}
\delta  = \frac{3c^2w^3w_2 + 2{M_P}^3w\z_0\left( w - 4w_2{\z_0}^2 \right) g }
  {2c^2w^2w_2\left( w + c^2{M_P}^2 w_2 + 6w_2{\z_0}^2 \right)  + 
    4{M_P}^3\z_0\left( w^2 + 2w_2\left( w - c^2{M_P}^2w_2 \right) {\z_0}^2 - 
       8{w_2}^2{\z_0}^4 \right) g }.
\nonumber \\
\end{eqnarray}
One can calculate vacuum expectation values of the $F$-terms:
\begin{eqnarray} \label{Fz}\nonumber 
&F^\z =&\frac{-2{\sqrt{2}}w{\left( \frac{\z_0}{{M_P}} \right) }^{\frac{3}{2}}}{{M_P}} + 
  \frac{2{\sqrt{2}}{\left( \frac{\z_0}{{M_P}} \right) }^{\frac{3}{2}}
     \left( w + c^2{M_P}^2w_2 + 4w_2{\z_0}^2 \right) \delta }{{M_P}}
\\ \nonumber 
&&+ 
  \frac{{2\sqrt{2}}\left( w - c^2{M_P}^2w_2 \right) {\z_0}^3{\sqrt{\frac{\z_0}{{M_P}}}}{\delta }^2}{c^2{M_P}^4}
\\ \nonumber 
&& + 
  \left( \frac{-5{\sqrt{2}}w{\left( \frac{\z_0}{{M_P}} \right) }^{\frac{3}{2}}}{{M_P}} + 
     \frac{{\sqrt{2}}{\left( \frac{\z_0}{{M_P}} \right) }^{\frac{3}{2}}
        \left( c^4{M_P}^4w_2 - 4w{\z_0}^2 + c^2{M_P}^2\left( w + 20w_2{\z_0}^2 \right)  \right) \delta }{c^2
        {M_P}^3} \right) \epsilon 
\\ \nonumber 
&& 
 + \frac{w{\left( \frac{\z_0}{{M_P}} \right) }^{\frac{3}{2}}\left( -7c^2{M_P}^2 + 8{\z_0}^2 \right) 
     {\epsilon }^2}{2{\sqrt{2}}c^2{M_P}^3}+ 
  \dots +\gamma \left( \frac{-16{\sqrt{2}}w{\z_0}^3{\sqrt{\frac{\z_0}{{M_P}}}}{\delta }^3}{c^2{M_P}^4}  \right.\\ \nonumber 
&& - 
    \frac{24{\sqrt{2}}w_2{\z_0}^3{\sqrt{\frac{\z_0}{{M_P}}}}{\delta }^3}{{M_P}^2}+ 
    \frac{48{\sqrt{2}}w{\z_0}^3{\sqrt{\frac{\z_0}{{M_P}}}}{\delta }^2\epsilon }{c^2{M_P}^4} + 
    \frac{48{\sqrt{2}}w_2{\z_0}^3{\sqrt{\frac{\z_0}{{M_P}}}}{\delta }^2\epsilon }{{M_P}^2}  
\\ \nonumber 
&& \left. 
   -  \frac{48{\sqrt{2}}w{\z_0}^3{\sqrt{\frac{\z_0}{{M_P}}}}\delta {\epsilon }^2}{c^2{M_P}^4}- 
    \frac{24{\sqrt{2}}w_2{\z_0}^3{\sqrt{\frac{\z_0}{{M_P}}}}\delta {\epsilon }^2}{{M_P}^2} + 
    \frac{16{\sqrt{2}}w{\z_0}^3{\sqrt{\frac{\z_0}{{M_P}}}}{\epsilon }^3}{c^2{M_P}^4} \right), \\
&& 
\end{eqnarray}

\begin{eqnarray}\label{Fs}\nonumber 
&F^S =&\frac{-2{\sqrt{2}}w{\left( \frac{\z_0}{{M_P}} \right) }^{\frac{3}{2}}}{{M_P}} + 
  \frac{8{\sqrt{2}}w_2{\z_0}^3{\sqrt{\frac{\z_0}{{M_P}}}}\delta }{{M_P}^2} + 
  \frac{{2\sqrt{2}}\left( w - c^2{M_P}^2w_2 \right) {\z_0}^3{\sqrt{\frac{\z_0}{{M_P}}}}{\delta }^2}{c^2{M_P}^4} 
\\ \nonumber 
&&+ 
  \left( \frac{-3{\sqrt{2}}w{\left( \frac{\z_0}{{M_P}} \right) }^{\frac{3}{2}}}{{M_P}} + 
     \frac{4{\sqrt{2}}\left( -w + 5c^2{M_P}^2w_2 \right) {\z_0}^3{\sqrt{\frac{\z_0}{{M_P}}}}\delta }{c^2{M_P}^4} \right) 
   \epsilon 
\\ \nonumber 
&& + \frac{w{\left( \frac{\z_0}{{M_P}} \right) }^{\frac{3}{2}}\left( -3c^2{M_P}^2 + 8{\z_0}^2 \right) {\epsilon }^2}
   {2{\sqrt{2}}c^2{M_P}^3} +
 \dots +\gamma \left( \frac{4{\sqrt{2}}w{\z_0}^5{\sqrt{\frac{\z_0}{{M_P}}}}{\delta }^4}{c^4{M_P}^6} +\right.\\ \nonumber 
&&  - 
    \frac{16{\sqrt{2}}w{\z_0}^5{\sqrt{\frac{\z_0}{{M_P}}}}{\delta }^3\epsilon }{c^4{M_P}^6}+ 
    \frac{24{\sqrt{2}}w{\z_0}^5{\sqrt{\frac{\z_0}{{M_P}}}}{\delta }^2{\epsilon }^2}{c^4{M_P}^6}\\ \nonumber 
&& 
\left. -  \frac{16{\sqrt{2}}w{\z_0}^5{\sqrt{\frac{\z_0}{{M_P}}}}\delta {\epsilon }^3}{c^4{M_P}^6} + 
    \frac{4{\sqrt{2}}w{\z_0}^5{\sqrt{\frac{\z_0}{{M_P}}}}{\epsilon }^4}{c^4{M_P}^6} \right). \\
\end{eqnarray}

When the superpotential depends on two independent combinations of moduli, 
the minimum lies at the point $\z  =\z_0(1+\delta_{split})$ and
$ S = \z $, where 
\begin{eqnarray} \label{delsp}
\delta_{split}\approx \nonumber
 \eta h\left( 9{w_2}^3{\z_0}^5
        {\left( c^3{M_P}^3 + 4c{M_P}{\z_0}^2 \right) }^2 + 
       2\,{\sqrt{3}}\left( 9c^2{M_P}^2 + 4{\z_0}^2 \right) 
        {\sqrt{{w_2}^6{\z_0}^{12}{\left( c^3{M_P}^3 + 4c{M_P}{\z_0}^2 \right) }^2}}
       \right)  \\ \nonumber
\left( 2{w_2}^2{\z_0}^4\left( 3c^2{M_P}^2 - 4{\z_0}^2 \right)  + 
       2^{\frac{1}{3}}{\left( 4{w_2}^3{\z_0}^7
             \left( 9c^2{M_P}^2 + 4{\z_0}^2 \right)  + 
            6{\sqrt{3}}{\sqrt{{w_2}^6{\z_0}^{12}
                 {\left( c^3{M_P}^3 + 4c{M_P}{\z_0}^2 \right) }^2}} \right) }^{\frac{2}{3}} \right) \\ \nonumber
  \Bigg/\Bigg(
     {24{w_2}^2{\z_0}^2{\left( c^2{M_P}^2 + 4{\z_0}^2 \right) }^2
     {\left( 2{w_2}^3{\z_0}^7\left( 9c^2{M_P}^2 + 4{\z_0}^2 \right)  + 
         3{\sqrt{3}}{\sqrt{{w_2}^6{\z_0}^{12}
              {\left( c^3{M_P}^3 + 4c{M_P}{\z_0}^2 \right) }^2}} \right) }^{\frac{4}{3}}} \Bigg).
 \nonumber\\ 
\end{eqnarray}


\begin{thebibliography}{350}
\bibitem{Krasnikov:jj}
N.~V.~Krasnikov,
Phys.\ Lett.\ B {\bf 193} (1987) 37.

\bibitem{Dixon:1990ds}
L.~J.~Dixon,
SLAC-PUB-5229,
{\it Talk given at 15th APS Div. of Particles and Fields General Mtg., Houston,TX, Jan 3-6, 1990}.

\bibitem{clmr} 
J.~A.~Casas , Z.~Lalak , C. ~Munoz and G.~G.~Ross,
Nucl.\ Phys.\  B {\bf 347} (1990) 243.

\bibitem{Aldazabal:1998mr}
G.~Aldazabal, A.~Font, L.~E.~Ibanez and G.~Violero,
Nucl.\ Phys.\ B {\bf 536} (1998) 29.
              
\bibitem{Itoyama:1985ay}
H.~Itoyama and J.~Leon,
Phys.\ Rev.\ Lett.\  {\bf 56} (1986) 2352.

\bibitem{Ibanez:1986xy}
L.~E.~Ibanez and H.~P.~Nilles,
Phys.\ Lett.\ B {\bf 169} (1986) 354.

\bibitem{Banks:1996ss}
T.~Banks and M.~Dine,
Nucl.\ Phys.\ B {\bf 479} (1996) 173.

\bibitem{Falkowski:2001sq}
A.~Falkowski, Z.~Lalak and S.~Pokorski,
Nucl.\ Phys.\ B {\bf 613} (2001) 189.

\bibitem{Ibanez:1999pw}
L.~E.~Ibanez, R.~Rabadan and A.~M.~Uranga,
Nucl.\ Phys.\ B {\bf 576} (2000) 285.

\bibitem{Lalak:1999ex}
Z.~Lalak, S.~Lavignac and H.~P.~Nilles,
Nucl.\ Phys. B {\bf 576} (2000) 399.

\bibitem{P} E. Poppitz, 
Nucl.\ Phys. B {\bf 542} (1999) 31.


\bibitem{Lalak:2001dv}
Z.~Lalak,
{\it Talk given at SUSY01,
Dubna, Russia, 11-17 Jun 2001}, arXiv: hep-th/0109074.

\bibitem{AbelServant:2001ab}
S.~A.~Abel, G.~Servant,
Nucl.\ Phys.\ B {\bf 597} (2001) 3; Nucl.\ Phys.\ B {\bf 611} (2001) 43.

\bibitem{Lalak:1999yq}
Z.~Lalak and S.~Thomas,
Nucl.\ Phys.\ B {\bf 575} (2000) 151.

\bibitem{Horava:1996ma}
P.~Horava and E.~Witten,
Nucl.\ Phys.\ B {\bf 475} (1996) 94.

\end{thebibliography}
\end{document}